\documentclass[EN,dvipsnames]{rcftex}

\usepackage{cite}
\usepackage{times}
\usepackage{array}
\usepackage{fancyhdr}
\usepackage{multicol}
\usepackage{xcolor}

\usepackage{anyfontsize}
\usepackage{notes2bib}
\usepackage[spanish]{babel} 

\usepackage{pgfplots}
\usepgfplotslibrary{groupplots}
\pgfplotsset{compat=1.18}

\usepackage[version=4]{mhchem}

\title{Transición Metal-Aislante descrita por la TFON}
\titleEN{Metal-Insulator Transition described by NOFT}

\author{J. F. H. Lew-Yee\toaff{a\emailto} and M. Piris\toaff{a,b,c\emailto}}

\affiliations{
\aff Donostia International Physics Center (DIPC), 20018 Donostia, Spain. felipe.lew.yee@dipc.org$^\emailto$
\aff Kimika Fakultatea, Euskal Herriko Unibertsitatea (UPV/EHU), 20018 Donostia, Spain.  mario.piris@ehu.eus$^\emailto$
\aff IKERBASQUE, Basque Foundation for Science, 48009 Bilbao, Spain.
}

\received{16/4/2025}
\approved{12/5/2025}

\abstractES{La transición metal–aislante (TMA) es un fenómeno fundamental en física de la materia condensada y una manifestación característica de las correlaciones electrónicas fuertes. Los sistemas basados en hidrógeno constituyen un modelo simple pero eficaz para estudiar la TMA, ya que su comportamiento aislante se debe exclusivamente a interacciones electrón–electrón. En este trabajo, investigamos cúmulos finitos de hidrógeno con geometría cúbica mediante la Teoría del Funcional de Orbitales Naturales (TFON), un método capaz de describir con precisión sistemas correlacionados más allá del campo medio. Nos centramos en dos indicadores clave de la TMA: la brecha de energía fundamental y el promedio armónico de la matriz densidad uniparticular atómica. Nuestros resultados muestran que la TFON reproduce con éxito la transición del estado aislante al metálico al disminuir la distancia interatómica. Al extrapolar la brecha al límite termodinámico, estimamos una distancia crítica $r_c\approx1.2$\AA, en excelente concordancia con estudios de Monte Carlo cuántico.}

\abstractEN{The metal–insulator transition (MIT) is a fundamental phenomenon in condensed matter physics and a hallmark of strong electronic correlations. Hydrogen-based systems offer a simple yet powerful model for investigating the MIT, as their insulating behavior arises purely from electron–electron interactions. In this work, we study finite hydrogen clusters with cubic geometries using Natural Orbital Functional Theory (NOFT), a method capable of accurately describing correlated systems beyond mean-field approaches. We focus on two key signatures of the MIT: the fundamental energy gap and the harmonic average of the atomic one-particle reduced density matrix. Our results show that NOFT captures the transition from insulating to metallic behavior as the interatomic distance decreases. By extrapolating the energy gap to the thermodynamic limit, we estimate a critical distance $r_c\approx1.2$\AA, in excellent agreement with quantum Monte Carlo benchmarks. These findings demonstrate the reliability of NOFT for describing strong correlation effects in large-scale models.}

\keyW{Metal–Insulator Transition, Strongly Correlated Electrons, Reduced Density Matrix Functional Theory, Natural Orbitals, Hydrogen Clusters}

\begin{document}

\maketitle

\section{Introduction}

The metal–insulator transition (MIT) \cite{Mott1968} is a central concept in condensed matter physics and one of the most compelling manifestations of electron correlation effects in solids. In these transitions, materials change their electronic phase from conducting to insulating or vice versa under the influence of external parameters such as temperature, pressure, doping level, or lattice strain. While the transition may appear continuous or abrupt depending on the system, it generally reflects a subtle interplay between electron localization and delocalization. These phenomena are not only of fundamental theoretical interest but also of increasing technological relevance due to their potential integration into future microelectronic, sensing, and switching devices \cite{Imada1998}. Materials exhibiting MIT behavior are therefore highly valued in the search for controllable, multifunctional components for next-generation electronics \cite{Georgescu2021,Lee2024}.

A historically significant and conceptually rich subclass of MIT materials are Mott insulators, whose study began in 1937 \cite{Boer1937} during a conference chaired by Sir Nevill Mott. At the time, transition-metal oxides such as NiO, MnO, and Fe$_2$O$_3$ were observed to behave as insulators despite having partially filled 3d electronic bands, which, according to conventional band theory, should result in metallic behavior. Building on the insights of de Boer, Verwey, and Peierls, Mott argued that this insulating behavior was caused by strong on-site Coulomb repulsion, which localized electrons and suppressed conduction even in the absence of a filled band. This challenged the established Bloch-Wilson picture and gave rise to the concept of the Mott transition, where electron–electron interactions, rather than band filling, govern the electronic phase of the system \cite{Brandow1976}.

Contemporary research continues to explore MITs across a wide range of materials and dimensionalities \cite{Kettemann2023,Bellomia2024}. In particular, two-dimensional crystals with specific lattice geometries have recently demonstrated that even moderate interactions can induce Mott insulating behavior \cite{Lowe2024}. These findings highlight the universality of the MIT in systems with diverse chemical compositions, dimensionalities, and electronic structures. They also underscore the need for accurate theoretical models capable of capturing the essential physics of strong correlation. In this context, idealized systems composed of hydrogen atoms under extreme conditions offer a unique platform to isolate and study the fundamental ingredients that govern MIT phenomena \cite{Ceperley1987, Morales2010, Silvera2010}.

Among idealized systems, a simple lattice of hydrogen atoms offers a particularly clear framework for illustrating the essential mechanism of a Mott transition in its most fundamental form \cite{McMinis2015, Motta2020, Landinez-Borda2024}. At large interatomic distances, each hydrogen atom behaves as an isolated unit, with one electron localized at each site. Exciting the system into a conducting state requires transferring an electron from its original site to a distant one, which is energetically penalized by the difference between the ionization potential and the electron affinity, known as the fundamental energy gap. This difference defines an effective Coulomb repulsion energy, usually denoted as U. As the lattice spacing decreases, the overlap between electronic wavefunctions increases, broadening the energy levels by a bandwidth W. The system becomes metallic when the kinetic energy gain associated with delocalization overcomes the Coulomb repulsion, which leads to a critical transition governed by the ratio U to W. This framework is now a cornerstone in the theoretical understanding of many-body systems that display correlated insulating and metallic phases.

While the hydrogen lattice provides an idealized but physically transparent model to explore the fundamental mechanism of the Mott MIT, accurately capturing its electronic properties remains a significant theoretical challenge. Traditional ground-state density functional theory \cite{Hohenberg1964, Kohn1965}, despite its success in weakly correlated systems, fails to describe the insulating behavior of strongly correlated materials in the absence of explicit symmetry breaking \cite{Liechtenstein1995}, as evidenced in well-known cases such as transition metal oxides. Even advanced many-body approaches like the GW approximation often do not predict insulating states unless long-range magnetic order is imposed \cite{Aryasetiawan1995,Rodl2009}. In contrast, methods that explicitly incorporate many-body correlation effects, such as dynamical mean field theory \cite{Kun2008,Miura2008} and natural orbital functional (NOF) theory (NOFT) \cite{Piris2007, Piris2024a}, have been shown to correctly capture insulating phases without invoking long-range spin order \cite{Sharma2013, Shinohara2015a}. This capability is particularly relevant in model hydrogen systems, where electron correlation alone drives localization. Within this context, NOFT offers a promising framework to study finite hydrogen clusters as minimal yet nontrivial systems for understanding the MIT. In the present work, we employ NOFT to investigate signatures of the MIT in hydrogen cubes, focusing on the fundamental energy gap and the harmonic average of the atomic one-particle reduced density matrix (1RDM) as key indicators, in order to highlight the role of electronic correlation beyond mean-field approximations.

This article is organized as follows. First, we introduce the key concepts of NOFT, along with the electron-pairing-based approximations used throughout this work. Next, the computational methodology is presented, followed by a discussion of the results. The article closes with a summary of the main findings and some final remarks.

\bigskip

\section{Natural Orbital Functional Theory}

NOFT is the formulation of the 1RDM functional theory \cite{Gilbert1975, Levy1979, Valone1980, Schilling2018} in the natural orbital (NO) representation \cite{Lowdin1955}. It offers a computationally efficient alternative \cite{Lew-Yee2021} to conventional wavefunction-based methods, which often exhibit steep scaling. By using the 1RDM as the central variable and reconstructing the two-particle reduced density matrix (2RDM) through well-founded approximations, NOFT allows for an accurate description of strongly correlated electronic states, showing particular robustness in multireference regimes \cite{Lopez2011, Ruiperez2013, Ramos-Cordoba2015, Lopez2019, Mitxelena2020a, Mitxelena2020b, Mitxelena2022, Mitxelena2024}. In this framework, the ground-state energy is expressed as a functional of the NOs and their occupation numbers (ONs), which are variationally optimized. The presence of significantly fractional ONs reveals a multiconfigurational character, a regime traditionally addressed by methods such as CASPT2 \cite{Pulay2011}. However, these approaches require active space selection and become computationally expensive as the number of correlated orbitals increases. In contrast, NOFT correlates all electrons in all orbitals within the chosen basis set, avoiding the need to define an active space and eliminating arbitrary user choices. This makes NOFT particularly suitable for processes such as bond breaking and formation \cite{Piris2024b}, where the optimal active space is not known a priori, while also offering greater accessibility for non-expert users.

In recent years, NOFT has undergone significant progress from both theoretical and computational perspectives \cite{Liebert2023a, Ding2024, Piris2024c}. A notable result of this progress is the family of Piris NOFs (PNOFs) \cite{Piris2013b, Piris2017, Piris2021a}, which have consistently demonstrated competitiveness with standard electronic structure methods. These functionals have shown effectiveness across a wide range of applications \cite{Franco2023, Lew-Yee2023a, Franco2025}, including the description of excited states \cite{Lew-Yee2024}, molecular dynamics simulations \cite{RiveroSantamaria2024}, and the mitigation of delocalization errors \cite{Lew-Yee2023b}. More recently, NOFs have been incorporated into quantum computing frameworks to enhance energy estimation efficiency within the variational quantum eigensolver (VQE) algorithm, resulting in the development of a NOF-VQE \cite{Lew-Yee2025a}. On the computational side, recent advances \cite{Piris2021, Franco2024, Lew-Yee2025b} have considerably reduced the cost of NOF calculations. A key improvement has been the integration of modern numerical techniques inspired by deep learning, particularly momentum-based optimization methods, which have significantly accelerated convergence \cite{Lew-Yee2025b}. These developments have enabled NOFT to treat strongly correlated systems efficiently, establishing it as a practical tool for large-scale applications.

The ground-state energy of a NOF is expressed in terms of the set of NOs $\{\phi_i\}$ and their ONs $\{n_i\}$ as
\begin{equation}
    E[\mathrm{N},\{n_i,\phi_i\}]= \sum _i n_i H_{ii} + \sum  _{ijkl} D[n_i,n_j,n_k,n_l] \langle ij|kl \rangle 
    \label{Enoft}   
\end{equation}
where the one- and two-electron integrals are given by
\begin{equation}
H_{ii} =\int d{\bf r} \phi _i ^*({\bf r})\left(-\frac{\nabla ^2 _r} {2}+v({\bf r}) \right) \phi _i ({\bf r})
\label{hpp}   
\end{equation}
\begin{equation}
    \langle ij|kl \rangle= \int \int d{\bf r}_1 d {\bf r}_2 \frac{\phi ^* _i ({\bf r}_1)\phi ^* _j ({\bf r}_2)\phi _k ({\bf r}_1)\phi _l({\bf r}_2)}{|{\bf r}_2 -{\bf r}_1|}
    \label{eri} 
\end{equation}
In Eq.~(\ref{hpp}), $v({\bf r})$ represents the nuclear potential derived from the molecular geometry within the Born–Oppenheimer approximation, assuming no external fields. The exact form of the electron–electron interaction energy functional is unknown, and different functional forms of $D[n_i,n_j,n_k,n_l]$ define distinct NOFs.

The approximate functional in Eq.~(\ref{Enoft}) explicitly depends on the 2RDM, requiring not only the N-representability of the 1RDM \cite{Coleman1963} but also that of the functional itself \cite{Piris2018d}. Specifically, the reconstructed $D[n_i,n_j,n_k,n_l]$ must satisfy the same N-representability conditions as an unreconstructed 2RDM \cite{Mazziotti2012}, in order to ensure the existence of a compatible N-electron system. Different reconstructions of $D$ under these constraints have given rise to the various PNOFs.

In this work, we focus on electron-pairing-based PNOFs \cite{Piris2018e}. Accordingly, we consider $\mathrm{N_{I}}$ unpaired electrons, which determine the system’s total spin $S$, while the remaining $\mathrm{N_{II}} = \mathrm{N - N_{I}}$ electrons form pairs with opposite spins, resulting in zero net spin contribution from the paired electrons. Within the spin-restricted formalism, all spatial orbitals $\{\varphi_{p}\}$ are doubly occupied in the ensemble, and the ONs of both spin components are equal \cite{Piris2019}. Following the partitioning of electrons into $\mathrm{N_{I}}$ and $\mathrm{N_{II}}$, the orbital space $\Omega$ is divided into two subspaces: $\Omega = \Omega_{\mathrm{I}} \oplus \Omega_{\mathrm{II}}$. The subspace $\Omega_{\mathrm{II}}$ consists of $\mathrm{N_{II}}/2$ mutually disjoint subspaces $\Omega_{g}$, each containing a reference orbital $\left|g\right\rangle$ for $g \leq \mathrm{N_{II}}/2$, along with $\mathrm{N}{g}$ associated orbitals $\left|p\right\rangle$ for $p>\mathrm{N_{II}}/2$. Taking spin into account, the total occupancy of each subspace $\Omega_{g}$ is equal to 2. Similarly, $\Omega_{\mathrm{I}}$ is composed of $\mathrm{N_{I}}$ mutually disjoint subspaces; however, each $\Omega_{g} \in \Omega_{\mathrm{I}}$ contains only a single orbital $g$ with $n_{g} = 1/2$, corresponding to one unpaired electron whose spin state remains unspecified. It follows that the trace of the 1RDM equals the total number of electrons N.

This study focuses on finite hydrogen clusters, for which PNOF7 \cite{Piris2017,Mitxelena2018a} is employed. Previous studies have demonstrated that PNOF7 accurately reproduces the potential energy curves of such systems in smaller-scale cases \cite{Mitxelena2020a,Mitxelena2020b}. In particular, it captures the correct physical behavior in both the bonding and dissociation regimes, yielding stable and reliable dissociation energies. 

The energy expression for PNOF7 is given by
\begin{equation}
E\left[\mathrm{N},\left\{ n_{p},\varphi_{p}\right\} \right] = E^\text{intra}+E^\text{inter} \label{pnof5}
\end{equation}
The intra-pair component is formed by summing the energies $E_{g}$ of electron pairs with opposite spins and the single-electron energies of unpaired electrons, specifically:
\begin{equation}
E^\text{intra} =\sum\limits _{g=1}^{\mathrm{N_{II}}/2}E_{g}+{\displaystyle \sum_{g=\mathrm{N_{II}}/2+1}^{\mathrm{N_{II}}/2+\mathrm{N_{I}}}}H_{gg}
\label{Eintra}
\end{equation}
\begin{equation}
E_{g} = 2 \sum\limits _{p\in\Omega_{g}}n_{p}H_{pp} + \sum\limits _{q,p\in\Omega_{g}} \Pi(n_q,n_p) L_{pq}
\end{equation}
Here, $L_{pq}=\left\langle pp|qq\right\rangle$ are the exchange-time-inversion integrals \cite{Piris1999}. The matrix elements $\Pi(n_q,n_p) = c(n_q)c(n_p)$, where $c(n_p)$ is defined by the square root of the ONs as follows:
\begin{equation}
    c(n_p) = \left.
  \begin{cases}
    \phantom{+}\sqrt{n_p}, & p \leq \mathrm{N_{II}}/2\\
    -\sqrt{n_p}, & p > \mathrm{N_{II}}/2 + \mathrm{N_{I}} \\
  \end{cases}
  \right. \>\>
  \label{eq:PNOF5-roots}
\end{equation}
The inter-subspace term is given by
\begin{equation}
E^\text{inter}=\sum\limits _{p,q=1}^{\mathrm{N}_B}\,'\,\{n_{q}n_{p}\left(2J_{pq}-K_{pq}\right)-\Phi_{q}\Phi_{p}L_{pq}\}\label{einter}
\end{equation}
where $J_{pq}=\left\langle pq|pq\right\rangle$ and $K_{pq}=\left\langle pq|qp\right\rangle$ are the Coulomb and exchange integrals, respectively. The term $\Phi_{p}=\sqrt{n_{p}h_{p}}$, with $h_{p}=1-n_{p}$, defines a correlation factor that becomes significant when the ON $n_{p}$ deviates from 0 or 1. $\mathrm{N}_{B}$ denotes the number of basis functions considered. The prime in the summation indicates that only inter-subspace terms are included. Notably, PNOF7 introduces inter-pair static electron correlation, as $\Phi_{p}$ increases in regions where orbitals exhibit strong multiconfigurational character \cite{Mitxelena2018a}.

\section{Methodology}

The transition from a metallic to an insulating phase is generally associated with either a structural reorganization or a mechanism driven by the electron correlation. In this work, we focus on the latter, which underlies the MITs observed in hydrogen clusters. As the interatomic distance increases, the overlap between atomic orbitals responsible for conduction decreases, resulting in a reduced bandwidth. In systems with strong electron correlation, this reduction can lead to a Mott transition, where Coulomb repulsion dominates over delocalization, ultimately opening an energy gap. A critical distance exists beyond which electronic overlap becomes insufficient to sustain metallic behavior, signaling the onset of insulating character. This concept forms the basis for a first estimation of the critical interatomic distance, $r_c$, derived from the balance between Coulomb interaction and electronic delocalization.

An estimate of $r_c$ can be obtained from fundamental physical arguments. The distance at which Coulomb interaction dominates over delocalization can be interpreted through the Heisenberg uncertainty principle ($\Delta p \Delta r \sim \hbar$). For an electron confined to a region of size $r_c$, the characteristic momentum is $\Delta p \sim \hbar / r_c$, yielding a kinetic energy of $E_k \sim (\Delta p)^2/2m \sim \hbar^2/2mr_c^2$, while the Coulomb potential energy between two electrons at distance $r_c$ is $E_p = e^2/4 \pi \epsilon_0 r_c$. The transition occurs when Coulomb repulsion becomes significantly larger than the kinetic energy, which we assume as $E_p = 10 \cdot E_k$. Solving this condition for $r_c$ gives $r_c$ = 5$a_0$ = 2.65 \AA. This value is close to the estimate obtained by assuming a homogeneous electron gas with a critical density $n_c \sim$ 0.01 $a_0^{-3}$; using the relation $n_c \sim r_c^{-3}$, one obtains $r_c \sim$ 4.64$a_0 \sim$ 2.45 \AA. However, studies on hydrogen chains and clusters show that the transition occurs at shorter distances. A diffusion quantum Monte Carlo study \cite{McMinis2015} benchmarks the transition between paramagnetic and antiferromagnetic body-centered cubic atomic hydrogen in its ground state, reporting a second-order metal–insulator transition at $r_c$ = 2.27$a_0$ = 1.2 \AA.

The most rigorous way to define the MIT is by identifying the opening of a gap in the density of states at the Fermi level. However, this criterion is not directly applicable to finite clusters, which inherently exhibit a discretized spectrum. Nevertheless, the transition can still be located as the point where the fundamental energy gap surpasses the effects of finite-size discretization. Since this discretization error is not known a priori, we adopt a finite-size scaling approach in which it is accounted for by extrapolating the fundamental gap, defined as $\delta = E(\mathrm{N} + 1) + E(\mathrm{N} - 1) - 2E(\mathrm{N})$, to the thermodynamic limit ($\mathrm{N} \rightarrow \infty$). To this end, we compute $\delta$ for hydrogen cubes of increasing size and for a set of interatomic distances $r$. The resulting $\delta(\mathrm{N})$ values are fitted as a function of N using a decaying polynomial model: 
\begin{equation}
    \delta(\mathrm{N}) = \delta_\infty + \frac{A}{\mathrm{N}^p}
    \label{poli}
\end{equation}
where $\delta_\infty$ is the extrapolated gap in the infinite system, $A$ and $p$ are fitting parameters, and N is the number of electrons in the system. The MIT is then identified as the point where $\delta_\infty$ becomes nonzero, indicating the emergence of an insulating phase in the thermodynamic limit.

This procedure enables a detailed analysis of the scaling behaviour of the gap: At small interatomic distances (metallic phase), the parameter $A$ is typically large, indicating strong finite-size effects, while $p$ is small, meaning that the gap decays slowly with increasing system size. In this regime, $\delta_\infty \approx 0$, and the observed gap is primarily due to finite-size discretization. At larger distances (insulating phase), $A$ tends to be smaller, and $p$ is larger, indicating that the gap converges more rapidly to a finite $\delta_\infty$. In this case, the extrapolated gap dominates, and finite-size effects become negligible. It should be noted that, since the fitting procedure is purely numerical, it can occasionally produce a slightly negative value of $\delta_\infty$. Although unphysical, this is interpreted as a numerical artifact resulting from the extrapolation overshooting a truly gapless spectrum. In such cases, the correct physical interpretation is that the system remains metallic.

\section{Results}

We begin our analysis by examining the symmetric dissociation of a large hydrogen cluster consisting of 512 atoms arranged in a cubic structure ($8\times8\times8$). All calculations were performed using the DoNOF software package \cite{Piris2021} with the 6-31G basis set \cite{Ditchfield1971} and the def2-universal-jkfit auxiliary basis set, employing real orbitals within a spin-restricted framework. The perfect pairing scheme ($\mathrm{N}{g}=1$) is used, as it is the highest pairing allowed by the negatively charged system. This leads to a fully correlated treatment of 512 electrons distributed over 512 spatial orbitals, making it a highly demanding electronic structure calculation. 

Each configuration was defined by uniformly setting the interatomic distance between the nearest neighbours, resulting in a series of hydrogen cubes that span from a delocalised metallic regime to a localised insulating one. In the dissociation limit, the system becomes a collection of 512 non-interacting hydrogen atoms. Importantly, the correlation effects in the hydrogen cube involve all electrons equally, posing a significant challenge for electronic structure methods. To contextualise our results, we refer to a representative set of Quantum Monte Carlo studies on hydrogen systems \cite{Silvera2010, McMinis2015, Motta2020, Landinez-Borda2024}, highlighting the broader relevance of our work within the field.

\begin{figure}
 \input{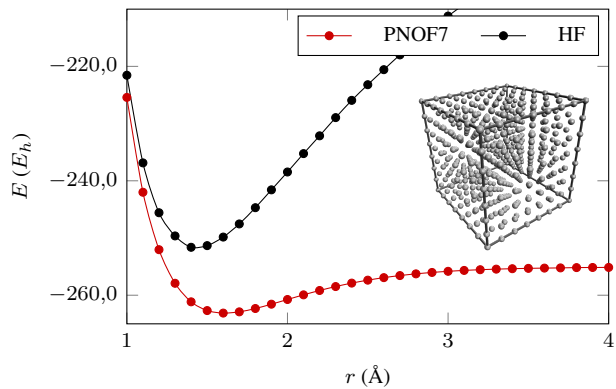}
 \caption{Potential energy curves for the symmetric dissociation of hydrogen cubes with 512 atoms.}
 \label{fig:H512}
\end{figure}

Fig. \ref{fig:H512} presents the total electronic energy as a function of the interatomic distance $r$ for hydrogen cubes with 512 atoms. The results obtained with PNOF7 (red) and restricted Hartree–Fock (HF, black) are shown for comparison. The PNOF7 curve exhibits a clear minimum around $r\approx1.6$ \AA. As the distance increases beyond this point, the energy gradually stabilizes, reflecting the transition toward the dissociated insulating state. In contrast, HF significantly overestimates the total energy across the entire range and fails to capture the energetic stabilization associated with electron correlation, particularly in the strongly correlated regime.

\bigskip

To quantitatively describe the MIT, we computed the harmonic average $\gamma$ of all off-diagonal elements $\Gamma^\text{ao}_{\mu\nu}$ of the 1RDM on the atomic orbital basis. In this regard, we generalized the definition on Ref.\cite{Sinitskiy2010} to count only interatomic elements as follow:
\begin{equation}
\gamma = \sqrt {\dfrac{1}{\mathrm{N}_{B}\left(\mathrm{N}_{B}-2\right)} \sum\limits_{\substack{\mu \in A,\; \nu \in B \\ A \ne B}} (\Gamma ^{ao}_{\mu\nu})^{2} }\label{eq:harmonic-average}
\end{equation}
where the summation indices indicate selecting the components associated with the atomic orbital $\phi_\mu$ on atom A and $\phi_\nu$ on atom B, while omitting the elements related to atomic orbitals on the same atom. Specifically, for the 6-31G basis set used here, which contains two atomic orbitals per hydrogen atom, this means excluding the $2 \times 2$ diagonal blocks. Consequently, $\mathrm{N}_{B}\left(\mathrm{N}_{B}-2\right)$ denotes the complete count of elements in the summation, although we note that the denominator depends on the number of atomic orbitals centered on each atom.

\bigskip

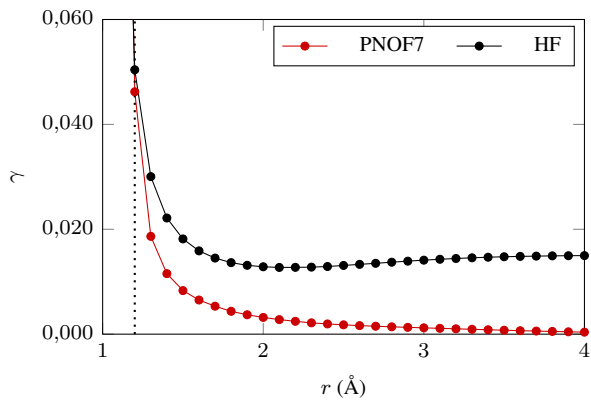
\begin{figure}
 \centering
 \begin{tikzpicture}
    \begin{axis}[
        xmin=1.0,
        xmax=4.0,
        ymax=0.06,
        ymin=0.000,
        xlabel= $r$ (\AA),
        ylabel= $\mathbf{\gamma}$,
        xtick = {1,2,3,4},
        y tick label style={
            /pgf/number format/.cd,
           fixed,
           fixed zerofill,
            precision=3,
            /tikz/.cd
        },
        scaled y ticks=false,
        legend style = { column sep = 10pt, legend columns = -1},
        width=0.9\hsize,
        height=0.65\hsize,
        font=\footnotesize,
    ]
    \addplot[red!80!black, mark=*, mark options={red!80!black}, mark size=1.5] table[x=r, y=h512] {h512-gamma.dat};
    \addlegendentry{PNOF7}
    \addplot[black, mark=*, mark options={black}, mark size=1.5] table[x=r, y=h512] {h512-gamma-HF.dat};
    \addlegendentry{HF}
    \addplot[
          dotted,
          thick
    ] coordinates {(1.2,0.0) (1.2,1.058)};
    \end{axis}
\end{tikzpicture}
 \caption{\label{fig:gamma} Harmonic average $\gamma$ of the off-diagonal elements of $\Gamma^\text{ao}$. The critical distance ($r_c$) for the metal-to-insulator transition is indicated by a vertical dotted line.}
\end{figure}
Fig.~\ref{fig:gamma} shows the evolution of this quantitative indicator of electronic delocalization. In the PNOF7 results (red), $\gamma$ decreases rapidly as the interatomic distance increases, signaling a transition from a delocalized metallic phase to a localized insulating phase. This behavior is consistent with the expected suppression of spatial coherence: as electrons localize, the off-diagonal elements $\Gamma^\text{ao}_{\mu\nu}$ tend toward zero, leading to a harmonic average that vanishes in the insulating limit. In contrast, HF results (black), exhibit a much slower decay and eventual saturation of $\gamma$, failing to capture the correlation-driven localization process.

\bigskip

\begin{figure}
 \begin{tikzpicture}
    \begin{axis}[
        xmin=1.8,
        xmax=4.0,
        ymax=0.65,
        ymin=0.0,
        xtick = {1,2,3,4},
        xlabel= $r$ (\AA),
        ylabel= $\mathbf{\delta}$,
        y tick label style={
            /pgf/number format/.cd,
            fixed,
            fixed zerofill,
            precision=1,
            /tikz/.cd
        },
        scaled y ticks=false,
        legend style = {column sep = 5pt, legend columns = 2, 
                        at={(1.0,0.0)}, anchor=south east},
        width=0.9\hsize,
        height=0.65\hsize,
        font=\footnotesize,
    ]

    \addplot[blue, mark=*, mark options={blue}, mark size=1.5, smooth] table[x=r, y=h8] {deltas.dat};
    \addlegendentry{\ce{H8}};
    \addplot[orange, mark=*, mark options={orange}, mark size=1.5, smooth] table[x=r, y=h64] {deltas.dat};
    \addlegendentry{\ce{H64}};
    \addplot[Green, mark=*, mark options={Green}, mark size=1.5, smooth] table[x=r, y=h216] {deltas.dat};
    \addlegendentry{\ce{H216}};
    \addplot[red!80!black, mark=*, mark options={red!80!black}, mark size=1.5, smooth] table[x=r, y=h512] {deltas.dat};
    \addlegendentry{\ce{H512}};
    \end{axis}
\end{tikzpicture}
 \caption{Fundamental energy gap for hydrogen cubes of various sizes.}
 \label{fig:deltas} 
\end{figure}
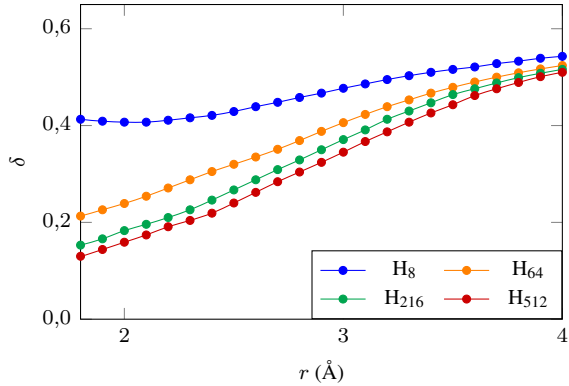

\begin{figure}
 \begin{tikzpicture}
\begin{axis}[
      xlabel={$\mathrm{N}$},
      ylabel={$\delta (\mathrm{N})$},
      ymax=0.6,
      ymin=0.1,
      xtick = {8,64,216,512},
      ytick = {0.1,0.2,0.3,0.4,0.5,0.6},
      scaled y ticks=false,
      width=0.85\hsize,
      height=0.7\hsize,
      font=\footnotesize,
      colorbar,
      colorbar style={xlabel=$r$ (\AA), ytick = {1,2,3,4}, font=\footnotesize,},
      colormap/viridis,
]

    \foreach \ylabel in {1.800, 1.900, 2.000, 2.100, 2.200, 2.300, 2.400, 2.500, 2.600, 2.700, 2.800, 2.900, 3.000, 3.100, 3.200, 3.300, 3.400, 3.500, 3.600, 3.700, 3.800, 3.900, 4.000} {
    \addplot [scatter, 
        mark=*,
        mark size=1.5,
        smooth,
        point meta=\ylabel,
        ] table [x=N, y=\ylabel] {fitting.dat};
    }

\end{axis}
\end{tikzpicture}
 \caption{Fundamental energy gap $\delta(\mathrm{N})$ as a function of system size N for hydrogen cubes at various interatomic distances $r$, color-coded as indicated by the color bar on the right. The data points correspond to calculated values, and the lines represent the fits using the decaying polynomial model of Eq.~\ref{poli}.}
 \label{fig:fitting}
\end{figure}
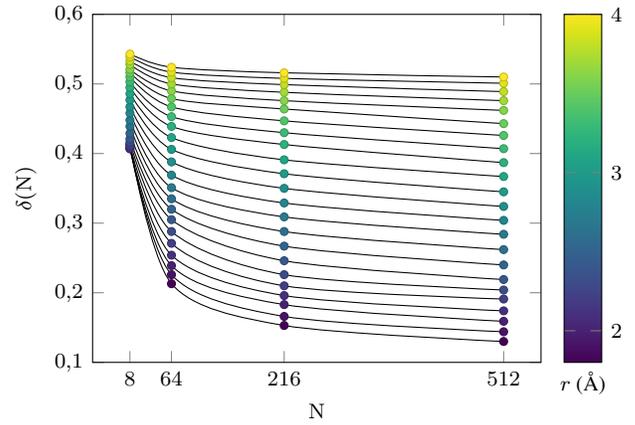

$r_c\approx1.2$ \AA \; is the estimated critical distance for the MIT, obtained through the finite-size scaling approach. The obtained fundamental energy gap $\delta(\mathrm{N})$ for hydrogen cubes of various sizes, ranging from $\mathrm{N}=8\;(2\times2\times2)$ to $\mathrm{N}=512\;(8\times8\times8)$, and for a set of interatomic distances $r$ are shown in Fig. \ref{fig:deltas}. 
The resulting values of $\delta(\mathrm{N})$ were fitted as a function of N using the decaying polynomial form of Eq.~\ref{poli}, as illustrated in Fig.~\ref{fig:fitting}. Each curve corresponds to a different interatomic distance $r$, color-coded from purple (short distances) to yellow (large distances). The fits demonstrate excellent agreement with the computed data points across all sizes and distances considered. For small $r$, the curves show finite-size dependence: $\delta(\mathrm{N})$ decreases significantly with increasing N, confirming that the finite gap observed in small clusters is a size effect. Conversely, at large interatomic distances (insulating regime), the gap is almost size-independent, and the extrapolated value $\delta_\infty$ closely matches the computed data for all cluster sizes. This behavior confirms the effectiveness of the extrapolation scheme and validates its use in identifying the metal–insulator transition from finite cluster data.

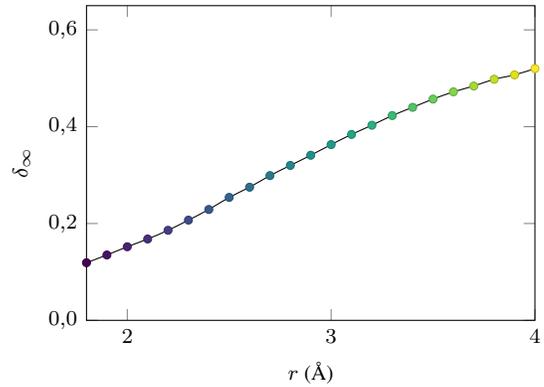
\begin{figure}
  \begin{tikzpicture}
\begin{axis}[
      xlabel= $r$ (\AA),
      ylabel={$\mathbf{\delta_\infty}$},
      ymax=0.65,
      ymin=0,
      xmin=1.8,
      xmax=4,
      xtick = {1,2,3,4},
      y tick label style={
          /pgf/number format/.cd,
          fixed,
          fixed zerofill,
          precision=1,
          /tikz/.cd
      },
      scaled y ticks=false,
      legend style = {column sep = 5pt, legend columns = 2, 
                      at={(1.0,0.0)}, anchor=south east},
      width=0.85\hsize,
      height=0.65\hsize,
      font=\footnotesize,
      colormap/viridis,
]
\addplot[
    mark=*,
    scatter,
    scatter src=explicit symbolic,
    point meta=\thisrow{r},
    mark size=1.5,
    smooth,
]
    table [x=r, y=delta, meta index=0] {extrapolated-delta.dat};
\end{axis}
\end{tikzpicture}
 \caption{Extrapolated fundamental energy gap $\delta_\infty$ as a function of the interatomic distance $r$, obtained from finite-size scaling of hydrogen cubes with up to 512 atoms.}
 \label{fig:delta_infinity}
\end{figure}

Fig.~\ref{fig:delta_infinity} shows the extrapolated values of the fundamental gap $\delta_\infty$ as a function of the interatomic distance $r$, obtained from finite-size scaling of hydrogen cubes. The curve exhibits a smooth and monotonic increase of $\delta_\infty$ with $r$, transitioning from metallic to insulating behavior. At short distances, $\delta_\infty$ follows an approximately linear trend. A linear regression in the interval $r \in$ [1.8 \AA, 2.5 \AA] yields the model $\delta_\infty = 0.190 \cdot r - 0.227$ with a coefficient of determination $R^2 = 0.993$. This fit predicts the closure of the gap at a critical distance $r=1.2$ \AA, corresponding to a critical electronic density of $n_c \approx 0.54 \; e^-/$\AA$^3$. Remarkably, this value agrees with diffusion Quantum Monte Carlo estimates of the Mott transition in atomic hydrogen \cite{McMinis2015}, confirming the robustness and accuracy of the present approach.

\bigskip \bigskip

\bigskip
\section{Conclusions}

We have presented a comprehensive analysis of the metal insulator transition (MIT) in finite hydrogen clusters using Natural Orbital Functional Theory (NOFT). By examining both the fundamental energy gap and the harmonic average of the atomic one-particle reduced density matrix, we assessed the onset of electron localization as a function of interatomic distance. The use of the electron-pair-based functional PNOF7 enabled an accurate description of both metallic and insulating regimes, and of the transition between them.

Through a finite-size scaling analysis applied to hydrogen cubes up to 512 atoms, we extrapolated the energy gap to the thermodynamic limit. From this analysis, we identified a critical interatomic distance of $r_c \approx 1.2$ \AA, in excellent agreement with diffusion Quantum Monte Carlo benchmarks. These results confirm that the MIT in hydrogen clusters arises from electron correlation effects and demonstrate the ability of NOFT to capture strong correlation phenomena in large systems with high accuracy and efficiency.

Altogether, the findings underscore NOFT’s potential as a reliable and efficient framework for exploring correlation-driven phase transitions in extended systems.

\bigskip

\section{Acknowledgments}

J. F. H. Lew-Yee acknowledges the DIPC and the MCIN program ``Severo Ochoa'' under reference AEI/CEX2018-000867-S for post-doctoral funding (Ref.: 2023/74.) M. Piris acknowledges funding from MCIN/AEI/10.13039/501100011033 (Ref.: PID2021-126714NB-I00) and the Eusko Jaurlaritza (Ref.: IT1584-22). The authors acknowledge the technical and human support provided by both the DIPC Supercomputing Center and SGIker (UPV/EHU, ERDF, EU).


\bigskip \bigskip

\section{References}

\begin{thebibliography}{67}
\providecommand{\url}[1]{\texttt{#1}}
\providecommand{\urlprefix}{URL }

\bibitem{Mott1968}
N.~F. Mott, Rev Mod Phys 40, 677 (1968).

\bibitem{Imada1998}
M.~Imada, A.~Fujimori and Y.~Tokura, Rev Mod Phys 70, 1039 (1998).

\bibitem{Georgescu2021}
A.~B. Georgescu, P.~Ren, A.~R. Toland, S.~Zhang, K.~D. Miller, D.~W. Apley, E.~A. Olivetti, N.~Wagner and J.~M. Rondinelli, Chem Mater 33, 5591 (2021).

\bibitem{Lee2024}
Y.~J. Lee, Y.~Kim, H.~Gim, K.~Hong and H.~W. Jang, Adv Mater 36, 2305353 (2024).

\bibitem{Boer1937}
J.~H. de~Boer and E.~J.~W. Verwey, Proc Phys Soc 49, 59 (1937).

\bibitem{Brandow1976}
B.~H. Brandow, Int J Quantum Chem Symp 10, 417 (1976).

\bibitem{Kettemann2023}
S.~Kettemann, Ann Phys 456, 169306 (2023).

\bibitem{Bellomia2024}
G.~Bellomia, C.~Mejuto-Zaera, M.~Capone and A.~Amaricci, Phys Rev B 109, 115104 (2024).

\bibitem{Lowe2024}
B.~Lowe, B.~Field, J.~Hellerstedt, J.~Ceddia, H.~L. Nourse, B.~J. Powell, N.~V. Medhekar and A.~Schiffrin, Nat Commun 15, 3559 (2024).

\bibitem{Ceperley1987}
D.~M. Ceperley and B.~J. Alder, Phys Rev B Condens Matter 36, 2092 (1987).

\bibitem{Morales2010}
M.~A. Morales, C.~Pierleoni, E.~Schwegler and D.~M. Ceperley, Proc Natl Acad Sci U S A 107, 12799 (2010).

\bibitem{Silvera2010}
I.~Silvera, Proc Natl Acad Sci USA 107, 12743 (2010).

\bibitem{McMinis2015}
J.~McMinis, M.~A. Morales, D.~M. Ceperley and J.~Kim, J Chem Phys 143, 194703 (2015).

\bibitem{Motta2020}
M.~Motta, C.~Genovese, F.~Ma, Z.-H. Cui, R.~Sawaya, G.~K.-L. Chan, N.~Chepiga, P.~Helms, C.~Jiménez-Hoyos, A.~J. Millis, U.~Ray, E.~Ronca, H.~Shi, S.~Sorella, E.~M. Stoudenmire, S.~R. White and S.~Zhang, Phys Rev X 10, 031058 (2020).

\bibitem{Landinez-Borda2024}
E.~J. Landinez-Borda, K.~O. Berard, A.~Lopez and B.~M. Rubenstein, Faraday Discuss 254, 500 (2024).

\bibitem{Hohenberg1964}
P.~Hohenberg and W.~Kohn, Phys Rev 136, B864 (1964).

\bibitem{Kohn1965}
W.~Kohn and L.~Sham, Phys Rev 140, A1133 (1965).

\bibitem{Liechtenstein1995}
A.~I. Liechtenstein, V.~I. Anisimov and J.~Zaanen, Phys Rev B 52, R5467 (1995).

\bibitem{Aryasetiawan1995}
F.~Aryasetiawan and O.~Gunnarsson, Phys Rev Lett 74, 3221 (1995).

\bibitem{Rodl2009}
C.~R\"odl, F.~Fuchs, J.~Furthm\"uller and F.~Bechstedt, Phys Rev B 79, 235114 (2009).

\bibitem{Kun2008}
J.~Kuně, A.~V. Lukoyanov, V.~I. Anisimov, R.~T. Scalettar and W.~E. Pickett, Nat Mater 7, 198 (2008).

\bibitem{Miura2008}
O.~Miura and T.~Fujiwara, Phys Rev B 77, 195124 (2008).

\bibitem{Piris2007}
M.~Piris, Adv Chem Phys 134, 387 (2007).

\bibitem{Piris2024a}
M.~Piris, Adv Quantum Chem 90, 15 (2024).

\bibitem{Sharma2013}
S.~Sharma, J.~K. Dewhurst, S.~Shallcross and E.~K.~U. Gross, Phys Rev Lett 110, 116403 (2013).

\bibitem{Shinohara2015a}
Y.~Shinohara, S.~Sharma, S.~Shallcross, N.~N. Lathiotakis and E.~K.~U. Gross, J Chem Theory Comp 11, 4895 (2015).

\bibitem{Gilbert1975}
T.~L. Gilbert, Phys Rev B 12, 2111 (1975).

\bibitem{Levy1979}
M.~Levy, Proc Natl Acad Sci USA 76, 6062 (1979).

\bibitem{Valone1980}
S.~M. Valone, J Chem Phys 73, 1344 (1980).

\bibitem{Schilling2018}
C.~Schilling, J Chem Phys 149, 231102 (2018).

\bibitem{Lowdin1955}
P.~O. L{\"{o}}wdin, Phys Rev 97, 1474 (1955).

\bibitem{Lew-Yee2021}
J.~F.~H. Lew-Yee, M.~Piris and J.~M. del Campo, J Chem Phys 154, 064102 (2021).

\bibitem{Lopez2011}
X.~Lopez, F.~Ruip{\'{e}}rez, M.~Piris, J.~M. Matxain and J.~M. Ugalde, ChemPhysChem 12, 1061 (2011).

\bibitem{Ruiperez2013}
F.~Ruip{\'{e}}rez, M.~Piris, J.~M. Ugalde and J.~M. Matxain, Phys Chem Chem Phys 15, 2055 (2013).

\bibitem{Ramos-Cordoba2015}
E.~Ramos-Cordoba, X.~Lopez, M.~Piris and E.~Matito, J Chem Phys 143, 164112 (2015).

\bibitem{Lopez2019}
X.~Lopez and M.~Piris, Theor Chem Acc 138, 89 (2019).

\bibitem{Mitxelena2020a}
I.~Mitxelena and M.~Piris, J Phys Condens Matter 32, 17LT01 (2020).

\bibitem{Mitxelena2020b}
I.~Mitxelena and M.~Piris, J Chem Phys 152, 064108 (2020).

\bibitem{Mitxelena2022}
I.~Mitxelena and M.~Piris, J Chem Phys 156, 214102 (2022).

\bibitem{Mitxelena2024}
I.~Mitxelena and M.~Piris, J Chem Phys 160, 204106 (2024).

\bibitem{Pulay2011}
P.~Pulay, Int J Quantum Chem 111, 3273 (2011).

\bibitem{Piris2024b}
M.~Piris, X.~Lopez and J.~M. Ugalde, J Phys Chem Lett 15, 12138 (2024).

\bibitem{Liebert2023a}
J.~Liebert and C.~Schilling, SciPost Phys 14, 120 (2023).

\bibitem{Ding2024}
L.~Ding, C.-L. Hong and C.~Schilling, Quantum 8, 1525 (2024).

\bibitem{Piris2024c}
M.~Piris, Chem Sci 15, 17284 (2024).

\bibitem{Piris2013b}
M.~Piris, Int J Quantum Chem 113, 620 (2013).

\bibitem{Piris2017}
M.~Piris, Phys Rev Lett 119, 063002 (2017).

\bibitem{Piris2021a}
M.~Piris, Phys Rev Lett 127, 233001 (2021).

\bibitem{Franco2023}
L.~Franco, J.~F.~H. Lew-Yee and J.~M. del Campo, AIP Advances 13, 065213 (2023).

\bibitem{Lew-Yee2023a}
J.~F.~H. Lew-Yee, J.~M. del Campo and M.~Piris, J Chem Theory Comput 19, 211 (2023).

\bibitem{Franco2025}
L.~Franco, R.~Rojas-Hernández, I.~A. Bonfil-Rivera, E.~Orgaz and J.~M. del Campo, Phys Chem Chem Phys  (2025).

\bibitem{Lew-Yee2024}
J.~F.~H. Lew-Yee, I.~A. Bonfil-Rivera, M.~Piris and J.~M. del Campo, J Chem Theory Comput 20, 2140 (2024).

\bibitem{RiveroSantamaria2024}
A.~Rivero-Santamaría and M.~Piris, J Chem Phys 160, 071102 (2024).

\bibitem{Lew-Yee2023b}
J.~F.~H. Lew-Yee, M.~Piris and J.~M. del Campo, J Chem Phys 158, 084110 (2023).

\bibitem{Lew-Yee2025a}
J.~F.~H. Lew-Yee and M.~Piris, J Chem Theory Comp 21, 2402 (2025).

\bibitem{Piris2021}
M.~Piris and I.~Mitxelena, Comput Phys Commun 259, 107651 (2021).

\bibitem{Franco2024}
L.~Franco, I.~A. Bonfil-Rivera, J.~F.~H. Lew-Yee, M.~Piris, J.~M.~del Campo and R.~A. Vargas-Hernández, J Chem Phys 160, 244107 (2024).

\bibitem{Lew-Yee2025b}
J.~F.~H. Lew-Yee, J.~M. del Campo and M.~Piris, Phys Rev Lett 134, 206401 (2025).

\bibitem{Coleman1963}
A.~J. Coleman, Rev Mod Phys 35, 668 (1963).

\bibitem{Piris2018d}
M.~Piris, G.~G.~N. Angilella and C.~Amovilli, eds., Many-body approaches Differ. scales a Tribut. to N. H. March Occas. his 90th Birthd., chapter~22, 261--278 (Springer, New York, 2018).

\bibitem{Mazziotti2012}
D.~A. Mazziotti, Chem Rev 112, 244 (2012).

\bibitem{Piris2018e}
M.~Piris, R.~Carb{\'{o}}-Dorca and T.~Chakraborty, eds., Quantum Chemistry at the Dawn of the 21st Century. Series: Innovations in Computational Chemistry, chapter~22, 593--620 (Apple Academic Press, 2018).

\bibitem{Piris2019}
M.~Piris, Phys Rev A 100, 32508 (2019).

\bibitem{Mitxelena2018a}
I.~Mitxelena, M.~Rodr{\'{i}}guez-Mayorga and M.~Piris, Eur Phys J B 91, 109 (2018).

\bibitem{Piris1999}
M.~Piris, J Math Chem 25, 47 (1999).

\bibitem{Ditchfield1971}
R.~Ditchfield, W.~J. Hehre and J.~A. Pople, J Chem Phys 54, 724 (1971).

\bibitem{Sinitskiy2010}
A.~V. Sinitskiy, L.~Greenman and D.~A. Mazziotti, J Chem Phys 133, 014104 (2010).

\end{thebibliography}

\end{document}